%% file: main.tex
\documentclass{article} %
\usepackage[final]{colm2026_conference}

\usepackage{microtype}
\usepackage{hyperref}
\usepackage{url}
\usepackage{booktabs}

\usepackage{lineno}

\input{packages.tex}
\input{macros.tex}

\input{plots/macros.tex}

\definecolor{darkblue}{rgb}{0, 0, 0.5}
\hypersetup{colorlinks=true, citecolor=darkblue, linkcolor=darkblue, urlcolor=darkblue}

\title{\textsc{Quasar}: A Programming Language Specialized for LLM Code Actions}

\author{%
\textbf{%
Stephen Mell\textsuperscript{1,2},
Botong Zhang\textsuperscript{2},
David Mell\textsuperscript{3},
Shuo Li\textsuperscript{2},
Ramya Ramalingam\textsuperscript{2},
} \\ \textbf{%
Nathan Yu\textsuperscript{3},
Stephan Zdancewic\textsuperscript{2},
Osbert Bastani\textsuperscript{2}%
} \\[0.5em]
\textsuperscript{1}Carnegie Mellon University \\
\textsuperscript{\phantom{1}}\texttt{stephenmell@cmu.edu} \\[0.3em]
\textsuperscript{2}University of Pennsylvania \\
\textsuperscript{\phantom{2}}\texttt{\{bzhang16,lishuo1,ramya23,stevez,obastani\}@engineering.upenn.edu} \\[0.3em]
\textsuperscript{3}Independent Researcher \\
\textsuperscript{\phantom{3}}\texttt{zraexy@gmail.com}, \texttt{nathany@alumni.cmu.edu}
}

\begin{document}

\ifcolmsubmission
\linenumbers
\fi

\maketitle

\input{body.tex}

\section*{Acknowledgments}
We thank the anonymous reviewers for their helpful feedback. This work was supported in part by NSF Awards CCF-1917852, CCF-2247088, CCF-2338777, and SLES 2331783, Amazon Research Award Fall 2023, and Amazon/ASSET Gift for Research in Trustworthy AI. Any opinions, findings, and conclusions or recommendations expressed in this material are those of the authors and do not necessarily reflect the views of funding entities.

\section*{Reproducibility Statement}
A software artifact~\citep{artifact} containing both the implementation and figure generation code is available at \url{https://github.com/stephenmell/quasar-colm2026-artifact}.

\bibliography{agent}
\bibliographystyle{colm2026_conference}

\appendix
\input{appendix-functionalization}

\end{document}

%% file: packages.tex
\usepackage[utf8]{inputenc} %
\usepackage[T1]{fontenc}    %
\usepackage{hyperref}       %
\usepackage{url}            %
\usepackage{booktabs}       %
\usepackage{amsfonts}       %
\usepackage{nicefrac}       %
\usepackage{microtype}      %
\usepackage{xcolor}         %

\usepackage{amsmath,amsthm,amssymb,bbm}
\usepackage{mathtools}
\usepackage{stmaryrd}       %
\usepackage{mathpartir}
\makeatletter
\newif\ifmpr@hasline
\def\mpr@extravskip{0.2em}
\def\mpr@linesep{%
   \ifx \mpr@vskip \empty \else
     \hrule height 0em depth \mpr@vskip width 0em
   \fi
   \ifmpr@hasline \ifx \mpr@extravskip \empty \else
     \hrule height 0em depth \mpr@extravskip width 0em
   \fi \fi
}
\def\mpr@htovlist{%
   \setbox \mpr@hlist
      \hbox {\strut
             \ifmpr@center \hskip -0.5\wd\mpr@hlist\fi
             \unhbox \mpr@hlist}%
   \ifdim \ht\mpr@vlist>\z@ \mpr@haslinetrue \else \mpr@haslinefalse \fi
   \setbox \mpr@vlist
      \vbox {\if@premisse
                \box \mpr@hlist
                \mpr@linesep
                \unvbox \mpr@vlist
             \else
                \unvbox \mpr@vlist
                \mpr@linesep
                \box \mpr@hlist
             \fi}%
}
\makeatother
\mprset{vskip=0.2em}
\usepackage{algorithm}
\usepackage[noend]{algpseudocode}
\makeatletter
\def\ALG@doentity{%
   \edef\ALG@thisblock{\csname ALG@currentblock@\theALG@nested\endcsname}%
   \expandafter\ifx\csname ALG@b@\ALG@L @\ALG@thisentity @\ALG@thisblock\endcsname\relax
      \def\ALG@thisblock{0}%
   \fi
   \ALG@getentitytext
   \ifnum\ALG@thisblock=0\else\ALG@doend\fi
   \ifx\ALG@text\ALG@x@notext
   \else
      \item
      \noindent\hskip\ALG@tlm
   \fi
   \expandafter\ifnum0=\csname ALG@b@\ALG@L @\ALG@thisentity @\ALG@thisblock\endcsname\else
      \ALG@dobegin
   \fi
   \def\ALG@entitiecommand{\ALG@displayentity}%
}
\makeatother
\usepackage{graphicx}
\usepackage{subcaption}
\usepackage[frozencache=true]{minted}

%% file: macros.tex
\newcommand{\toolname}{\textsc{Quasar}}
\newcommand{\toolnameabrev}{\toolname{}: QUick And Secure And Reliable}

\newmintinline[pyinline]{python}{}
\newcommand{\pyinlinem}[1]{\mbox{\pyinline{#1}}}

%% file: plots/macros.tex
\newcommand{\GqaTaskCount}{1000}
\newcommand{\AdTaskCount}{93}
\newcommand{\BcpSubsetSize}{80}

\newcommand{\ModelCodegen}{GPT-5}
\newcommand{\ModelSftTeacher}{GPT-5}
\newcommand{\BcpModel}{GPT-5-nano}
\newcommand{\ModelCodegenEffort}{Medium}
\newcommand{\ModelSftStudent}{GPT-4.1-nano}

\newcommand{\SpeedupRuntimeFloor}{100\,\mathrm{ms}}
\newcommand{\GqaSpeedupOverall}{$18\% \pm 23$}
\newcommand{\GqaSpeedupFracImprovable}{$49\%$}
\newcommand{\GqaSpeedupImprovable}{$37\% \pm 19$}
\newcommand{\AdSpeedupOverall}{$27\% \pm 32$}
\newcommand{\AdSpeedupFracImprovable}{$38\%$}
\newcommand{\AdSpeedupImprovable}{$54\% \pm 22$}
\newcommand{\BcpSpeedupOverall}{$93\% \pm 4$}

\newcommand{\GqaRoundsOverall}{$26\% \pm 29$}
\newcommand{\GqaRoundsFracImprovable}{$50\%$}
\newcommand{\GqaRoundsImprovable}{$53\% \pm 17$}
\newcommand{\AdRoundsOverall}{$26\% \pm 26$}
\newcommand{\AdRoundsFracImprovable}{$57\%$}
\newcommand{\AdRoundsImprovable}{$44\% \pm 19$}
\newcommand{\BcpRoundsOverall}{$97\% \pm 1$}

\newcommand{\ConfTargetError}{$10\%$}
\newcommand{\ConfNumSplits}{100}
\newcommand{\GqaConfError}{$7.9\% \pm 2.8$}
\newcommand{\GqaConfUncertain}{$58.2\% \pm 8.3$}
\newcommand{\AdConfError}{$8.6\% \pm 12.8$}
\newcommand{\AdConfUncertain}{$84.1\% \pm 23.8$}

\newcommand{\SftTrainSize}{900}
\newcommand{\SftHeldOutSize}{100}

%% file: body.tex
\begin{abstract}
Large language models (LLMs) often call external tools to solve tasks. One effective strategy is for LLMs to write code, enabling them to use complex control flow such as conditionals and loops. Such \emph{code actions} are typically represented as Python code, since LLMs are proficient at writing it. However, many programming language features that would support more effective code actions are difficult to implement for Python. We propose separating \emph{internal code} that captures program logic from \emph{external calls} to tools that interact with the world. New features can then easily be implemented by (1) annotating external calls with the effects relevant to that feature, and (2) modifying the execution of the internal code to track this information. We develop a novel programming language, \textsc{Quasar}, that implements this idea. To illustrate its utility, we implement several useful features on top of \textsc{Quasar} to enhance code actions: access control with batched user queries to improve security, autoparallelization of external calls to reduce latency, and conformal prediction for uncertainty quantification to mitigate hallucinations.
\end{abstract}

\section{Introduction}

Large language models (LLMs) have demonstrated remarkable reasoning capabilities. A powerful way to augment these capabilities is via \emph{tool calls}~\citep{schick2023toolformer}, which let the LLM write function calls whose results are appended to the context before generation continues. Tools include functions to read and edit files~\citep{yang2024sweagentagentcomputerinterfacesenable}, query knowledge sources such as databases~\citep{lewis2020retrievalaugmentedgenerationknowledgeintensivenlp}, store information in external memory across interactions~\citep{liu2024llmconversationalagentmemory, maharana2024evaluatinglongtermconversationalmemory}, and access user input/output devices such as the mouse, keyboard, and screen~\citep{computeruse}. Modern approaches often combine tool calling with chain-of-thought reasoning~\citep{wei2022chainofthoughtpromptingelicitsreasoning} to create autonomous \emph{agents} capable of solving problems through sequential tool calls~\citep{yao2023reactsynergizingreasoningacting}.

Recently, there has been a shift toward providing tools in the form of software APIs and having LLMs write code that invokes them, an approach known as \emph{code actions} or \emph{programmatic tool calling}. This strategy enables the LLM to write code that includes control flow to facilitate more complex interactions, such as automating iterative tasks by writing loops, without the result of every tool call needing to pass through the LLM context. Code actions have been shown to provide better task performance, have lower latency, and use fewer tokens than traditional tool calling~\citep{wang2024executablecodeactionselicit,ptc,zhang2025recursivelanguagemodels}, and even to provide protection against prompt-injection attacks~\citep{camel}.

Existing systems predominantly use Python~\citep{wang2024executablecodeactionselicit} and Bash~\citep{claudecode, yang2024sweagentagentcomputerinterfacesenable} for code actions, since the prevalence of these languages makes it easy for LLMs to write code in them~\citep{cassano2023multiple, orlanski2023measuring}. However, these kinds of general-purpose languages were not designed with code actions in mind and lack potentially useful features---for instance, access control restrictions (for calls to irreversible APIs that might require human approval) to improve security, autoparallelization to improve latency, and uncertainty quantification (for code actions calling other models, e.g., sub-agents) to improve reliability.

\begin{figure}[t]
\centering
\hfill
\begin{subfigure}{0.49\textwidth}
\centering
\includegraphics[width=0.75\textwidth]{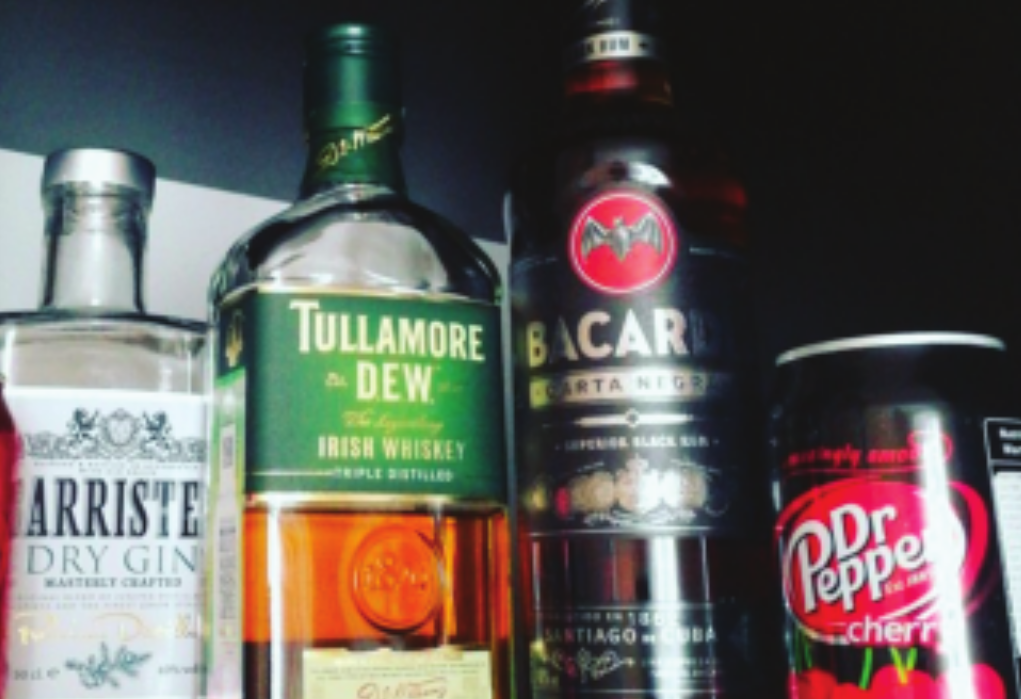}
\vspace{0.25em} \\
{\small Is there an alcoholic drink in this image?}
\caption{}
\end{subfigure}
\hfill
\begin{subfigure}{0.49\textwidth}
\begin{minted}[fontsize=\small]{python3}
drink_patches = image_patch.find("drink")
found = False
for drink_patch in drink_patches:
    if drink_patch.simple_query(
        "Does this have alcohol?"
    ) == "yes":
        found = True
return found
\end{minted}
\vspace{1.5em}
\caption{}
\end{subfigure}
\caption{Illustrative example~\citep{surís2023vipergptvisualinferencepython} of an image and a natural language question about that image (a), and a code action answering that question (b).}
\label{fig:example}
\end{figure}

Furthermore, implementing these features directly in general-purpose languages is complicated by their design. Specifically, we observe that code actions combine two distinct kinds of computation: \emph{internal code}, which performs complex logic but has no effect on the world (e.g., data manipulation or control flow), and \emph{external calls}, which affect the world (e.g., sending emails or invoking other models). Consider the example in Figure~\ref{fig:example}, where the program invokes sub-agents (\pyinline{find} and \pyinline{simple_query}) to answer the question. The \pyinline{for} loop, \pyinline{if} statement, and computation of \pyinline{found} are the internal computations, which orchestrate the external calls to sub-agents.

General-purpose languages such as Python and Bash do not separate external calls from internal code. Yet, doing so can make it significantly easier to implement desirable language features. At a high level, annotations document the relevant effects of external calls (e.g., resources accesses, data dependencies, or prediction uncertainties), and internal code tracks these effects during execution. In our example,
suppose we want to apply conformal prediction~\citep{vovk2005algorithmic} to quantify uncertainty. The external calls to sub-agents are modified to return sets of labels, and the execution of the internal code is modified to operate on these sets (e.g., Booleans are now \pyinline{True}, \pyinline{False}, or $\{\pyinlinem{True},\pyinlinem{False}\}$). This approach can also be applied to access control, where a key challenge is approval fatigue~\citep{beautement2008compliance, automode}. Traditional languages block execution until actions are approved by the user, forcing sequential approvals. The number of user interactions can be reduced by presenting actions for approval in large batches. To realize this approach, external calls can be annotated with the resources they access, and internal code can be ``opportunistically'' executed~\citep{opal} to batch as many external calls as possible. Users are only asked for approval when no further progress can be made without it. Opportunistic execution also enables autoparallelization, where external calls are labeled with data dependencies and internal code executes external calls in parallel when possible. To some degree, these features can be implemented in existing languages via ``tracing'', executing code to build computation graphs~\citep{tracing}. However, tracing-based approaches rely on the host language for control flow and thus do not capture it. This limits the kinds of features that can be supported, e.g., preventing conformal prediction from executing both branches of an \pyinline{if} statement when the condition is $\{\pyinlinem{True},\pyinlinem{False}\}$.

We propose a new language called \toolname{}\footnote{\toolnameabrev{}} that explicitly separates internal code from external calls and serves as a platform where features supporting code actions are simple to implement. \toolname{} can support arbitrary Python libraries by treating them as external tool calls. To enable LLMs to generate \toolname{} code, we rely on recent work compiling an expressive subset of Python into $\lambda$ calculus~\citep{poppy}, and restrict the LLM to generating code within this subset. To demonstrate the utility of \toolname{}, we implement and evaluate three diverse features: batched resource approval for improved access control; autoparallelization of external calls; and conformal semantics for uncertainty quantification.

In summary, we make the following contributions:
\begin{itemize}
\item We propose distinguishing between internal computation and external tool calls in code actions, and we design a language specialized for this.
\item We implement this idea as \toolname{}, a flexible platform for alternative program semantics and effect annotations.
\item We show its extensibility by implementing and evaluating three diverse features: access control approval batching, autoparallelization, and conformal prediction.
\item We show how to generate \toolname{} code from LLMs without reducing performance on downstream tasks.
\end{itemize}

\textbf{Related work.} Numerous studies have explored augmenting LLMs with code actions to improve task performance, including CodeAct~\citep{wang2024executablecodeactionselicit}, ViperGPT~\citep{surís2023vipergptvisualinferencepython}, and AppWorld~\citep{trivedi2024appworldcontrollableworldapps}. Despite these advancements, several challenges remain, including security and privacy risks~\citep{he2026emergedsecurityprivacyllm, andriushchenko2025agentharmbenchmarkmeasuringharmfulness}, persistent hallucination issues~\citep{liu2024uncertaintyestimationquantificationllms, li2024redoexecutionfreeruntimeerror}, and concerns over computational efficiency~\citep{yao2023treethoughtsdeliberateproblem}. Recent work has proposed leveraging code actions to improve security, where LLM-generated code can be analyzed for security vulnerabilities~\citep{camel}. However, it does not support batching of user approval requests, and it is not extensible (e.g., for latency or reliability improvements). Other work has addressed conformal prediction of functional programs with neural components~\citep{ramalingam2024} and automatic parallelization of functional programs~\citep{opal}. Though we draw on insights from this work, neither considers programs generated by LLMs nor imperative features of languages such as Python. Finally, \toolname{} relies on the compiler proposed in \citet{poppy}; that work focuses exclusively on autoparallelization, does not provide the broader benefits of \toolname{} for implementing code actions, and does not show how to generate programs in its supported subset of Python.

\section{\toolname{} programming language}
\label{sec:lang}
We first describe the syntax and semantics of \toolname{} programs~(\S\ref{sec:lang:syntaxsemantics});
then, we describe how \toolname{} code is generated by LLMs~(\S\ref{sec:lang:generation}).
We show a running example in Figure~\ref{fig:main} for the problem in Figure~\ref{fig:example}.

\begin{figure}[t]
\centering\scriptsize
\begin{tabular}{@{}l@{\quad}|@{\quad}l@{}}
\toprule\noalign{\vspace{0.5em}}
$\begin{aligned}
P_1&=\begin{array}{l}
\pyinlinem{drink_patches = image_patch.find("drink")} \\
\pyinlinem{found = False} \\
\pyinlinem{for drink_patch in drink_patches:} \\
\qquad\pyinlinem{if drink_patch.simple_query(} \\
\qquad\qquad\pyinlinem{"Does this have alcohol?"} \\
\qquad\pyinlinem{) == "yes":} \\
\qquad\qquad\pyinlinem{found = True} \\
\pyinlinem{return found}
\end{array}
\end{aligned}$
&
$\begin{aligned}
E_1&=\{(\pyinlinem{image_patch.find("drink")},\varnothing)\} \qquad\rightsquigarrow \\
E_2&= \{(\pyinlinem{image_patch.find("drink")},\pyinlinem{[patch1, patch2]})\}
\end{aligned}$ \\ \noalign{\vspace{0.5em}}\midrule\noalign{\vspace{0.5em}}
$\begin{aligned}
P_2&=\begin{array}{l}
\pyinlinem{found = False} \\
\pyinlinem{for drink_patch in [patch1, patch2]:} \\
\qquad\pyinlinem{if drink_patch.simple_query(} \\
\qquad\qquad\pyinlinem{"Does this have alcohol?"} \\
\qquad\pyinlinem{) == "yes":} \\
\qquad\qquad\pyinlinem{found = True} \\
\pyinlinem{return found}
\end{array}
\end{aligned}$
&
$\begin{aligned}
P_3&=\begin{array}{l}
\pyinlinem{found = False} \\
\pyinlinem{if patch1.simple_query(} \\
\qquad\qquad\pyinlinem{"Does this have alcohol?") == "yes":} \\
\qquad\pyinlinem{found = True} \\
\pyinlinem{if patch2.simple_query(} \\
\qquad\qquad\pyinlinem{"Does this have alcohol?") == "yes":} \\
\qquad\pyinlinem{found = True} \\
\pyinlinem{return found}
\end{array}
\end{aligned}$ \\ \noalign{\vspace{0.5em}}\midrule\noalign{\vspace{0.5em}}
$\begin{aligned}
E_3&=\{(\pyinlinem{patch1.simple_query(...)},\varnothing), \\
&\qquad\qquad(\pyinlinem{patch2.simple_query(...)},\varnothing)\} \qquad \rightsquigarrow \\
E_4&=\{(\pyinlinem{patch1.simple_query(...)},\pyinlinem{"no"}), \\
&\qquad\qquad(\pyinlinem{patch2.simple_query(...)},\pyinlinem{"yes"})\}
\end{aligned}$
&
$\begin{aligned}
P_4&=\begin{array}{l}
\pyinlinem{found = False} \\
\pyinlinem{if "no" == "yes":} \\
\qquad\pyinlinem{found = True} \\
\pyinlinem{if "yes" == "yes":} \\
\qquad\pyinlinem{found = True} \\
\pyinlinem{return found}
\end{array}
\end{aligned}$ \\ \noalign{\vspace{0.5em}}\midrule\noalign{\vspace{0.5em}}
$\begin{aligned}
P_5&=\pyinlinem{return True}
\end{aligned}$
& \\ \noalign{\vspace{0.5em}}
\bottomrule
\end{tabular}
\caption{Given program $P_1$ from Figure~\ref{fig:example}, one way \toolname{} could execute it is as follows. First, it dispatches the external call \pyinline{image_patch.find("drink")}, resulting in execution set $E_1$. Eventually, this external call finishes running and returns \pyinline{[patch1, patch2]}, resulting in execution set $E_2$, after which \toolname{} applies $R_{\text{ext}}$ to substitute this value into $P_1$ to obtain $P_2$. Then, \toolname{} applies an internal rule to unroll the for loop in $P_2$ to obtain $P_3$. Next, it dispatches both \pyinline{patch1.simple_query(...)} and \pyinline{patch2.simple_query(...)} resulting in execution set $E_3$. As before, these external calls finish running and return \pyinline{"no"} and \pyinline{"yes"}, respectively, yielding $E_4$, so \toolname{} applies $R_{\text{ext}}$ twice (once for each external call) to substitute these values into $P_3$ to obtain $P_4$. Finally, \toolname{} applies additional internal rules to simplify the conditionals in $P_4$, resulting in terminal program $P_5$.
}
\label{fig:main}
\end{figure}

\subsection{Syntax and semantics}
\label{sec:lang:syntaxsemantics}

A \toolname{} program $P\in\mathcal{P}$ consists of standard syntactic constructs such as conditionals, loops, and function calls. The execution of a \toolname{} program $P\in\mathcal{P}$ is expressed as a set of \emph{rewrite rules} $\mathcal{R}$. If a rule $R\in\mathcal{R}$ is applicable to $P$, then it transforms $P$ into a new program $P'$, which we denote by $P\rightarrow_R P'$. In general, there may be multiple possible programs $P'$ satisfying $P\rightarrow_R P'$, such as if the rule $R$ is applicable to different parts of $P$. If there is any rewrite rule $R$ mapping $P$ to $P'$, then we simply write $P\rightarrow P'$. We give the full set of rewrite rules in Figure~\ref{fig:fulllang:rules} in Appendix~\ref{sec:fulllang}. The core semantics does not require that these rules be applied in a particular order; regardless of the order, the program result will always be the same (a property known as \emph{confluence}). Thus, \toolname{} extensions are free to choose whatever order is most appropriate.

There are two kinds of rewrite rules: internal rules $\mathcal{R}_{\text{int}}$ and external rules $\mathcal{R}_{\text{ext}}$. Internal rules do not have \emph{effects}, meaning they do not have consequences external to the program, including network calls, system calls, calls to external APIs, or even printing. Internal rules perform transformations such as substituting variables, unrolling loops, and resolving conditionals; these rules are applicable when the necessary values are constants (e.g., a conditional whose predicate is \pyinline{True} or \pyinline{False}).

For example, in program $P_2$ in Figure~\ref{fig:main}, the list in the for loop is a constant value \pyinline{[patch1, patch2]}, so \toolname{} applies a rule to unroll the for loop, resulting in $P_3$. Similarly, in $P_4$, it can apply rewrite rules to transform the predicate \pyinline{"no" == "yes"} into \pyinline{False} and the predicate \pyinline{"yes" == "yes"} into \pyinline{True}, after which it can rewrite the conditionals to obtain $P_5$.

There is only one external rule $\mathcal{R}_{\text{ext}}=\{R_{\text{ext}}\}$. This rule is designed to enable calls to \emph{external functions} $f\in\mathcal{F}_{\text{ext}}$. Unlike a typical function, which is implemented as \toolname{} code, an external function is implemented in Python; thus, external functions can perform desirable effects such as printing a value or calling an LLM to obtain its output. An \emph{external call} in program $P$ is a statement $S=y\gets f(x_1,...,x_k)$ that calls an external function $f\in\mathcal{F}_{\text{ext}}$.

\toolname{} may execute external calls as soon as all their arguments are available. Formally, an external call $S=y\gets f(x_1,...,x_k)$ in a program $P$ is \emph{dispatchable} if all of $x_1,...,x_k$ are values (e.g., \pyinline{0}, \pyinline{True}, or \pyinline{"foo"}; recall that variables become values as the program is incrementally rewritten). As \toolname{} performs rewrites, it keeps track of the currently executing external calls $(S,B)\in E$, where $S$ is a pointer to the external call in the current program $P$ (preserved by rewrites) and $B$ is a pointer to a value that is initially $\varnothing$ but is eventually set to the output of the external function. After a rewrite $P\rightarrow P'$, \toolname{} identifies all the dispatchable external calls $S$ in $P'$ that are not yet in $E$; for each $S=y\gets f(x_1,...,x_k)$, it executes $f$ on $x_1,...,x_k$ in a separate thread $T$, and adds the pending call $(S,B)$ to $E$. The thread $T$ is also given $B$; once it finishes executing the external function $f$, it writes the output of $f$ to $B$ and terminates. Then, \toolname{} applies the rewrite rule $R_{\text{ext}}$ to the current program (which may no longer be $P'$) to substitute the value in $B$ into the program.

For example, in Figure~\ref{fig:main}, given the initial program $P_1$, \toolname{} executing with the parallel semantics immediately dispatches the external call \pyinline{image_patch.find("drink")} in a separate thread, leading to execution set $E_1$. When this thread finishes, it writes the result \pyinline{[patch1, patch2]} to $B$, resulting in $E_2$. This allows $R_{\text{ext}}$ to be applied to $P_1$, obtaining $P_2$. Similarly, as soon as \toolname{} rewrites $P_2$ to $P_3$, it dispatches two external calls \pyinline{patch1.simple_query(...)} and \pyinline{patch2.simple_query(...)}, resulting in $E_3$; these execute and return \pyinline{"no"} and \pyinline{"yes"}, respectively, resulting in $E_4$. Finally, \toolname{} applies $R_{\text{ext}}$ twice to substitute these values into $P_3$, resulting in $P_4$. %
A program $P$ is \emph{terminal} (i.e., finished executing) if no rules are applicable to $P$ and there are no pending external calls. $P_4$ is then rewritten to $P_5$, which is terminal.

\subsection{Generating \toolname{} code}
\label{sec:lang:generation}

\begin{figure}
  \hfill
  \begin{subfigure}[b]{0.3\textwidth}
    \tiny
    \centering
\begin{verbatim}
({77: '.find', 83: '.simple_query'},
 ('def',
  75,
  ((76,),
   ((('prim', 78, 'drink'),
     ('call', (79,), 77, (76, 78)),
     ...
         ('call', (91,), 0, (85,)),
         ('call', (92,), 91, (86, 88)),
         ('call', (90,), 92, ())),
        (90,)))),
     ('call', (93,), 79, (80, 81))),
    (93,)))))
\end{verbatim}
    \caption{}
    \label{fig:quasarcode:quasar}
  \end{subfigure}
  \hfill
  \begin{subfigure}[b]{0.45\textwidth}
\small
\centering
\input{plots/eval_generation.tex}
\caption{}
\label{tab:eval:generation}
  \end{subfigure}
  \hfill\,
\caption{The Python code from Figure~\ref{fig:example} as \toolname{} code (a). Comparison of different code generation approaches on the GQA and AgentDojo (AD) datasets (b). ``Execution'' is the fraction of generated programs that execute successfully (i.e., no syntax or runtime errors). ``Accuracy'' is the fraction of generated programs that output the ground truth label.}
\end{figure}

Though, for illustration, we have written example code with a syntax similar to Python, raw \toolname{} code actually looks very different. A key challenge is that LLMs have never seen it before, and 
we find that they struggle to generate it directly. Instead, our strategy is to have the LLM generate Python and then \emph{transpile} this Python code to \toolname{} using an existing compiler~\citep{poppy}. That is, the LLM generates the code in Figure~\ref{fig:example}, we transpile it to the code in Figure~\ref{fig:quasarcode:quasar}, and then the \toolname{} interpreter executes it. It is very challenging to transpile unrestricted Python to \toolname{}, since doing so would inherit all the challenges of
applying alternative semantics to Python.
However, many practical code actions do not use the unsupported language features of Python (e.g., classes and inheritance); intuitively, LLMs are trying to perform actions, not write complex software.
Thus, our transpiler supports a restricted subset of Python carefully chosen to balance expressiveness and ease of transpilation; we provide details on this transpilation in Appendix~\ref{sec:transpilation}. We consider three strategies for generating Python code in this subset.
(1) \emph{Instruction:} It is often sufficient to instruct the LLM to do so in the system prompt.
(2) \emph{Multi-turn:} It is also possible to feed error messages from our transpiler back to the LLM in a multi-turn loop. Importantly, this feedback is based only on static program information; in contrast, traditional multi-turn feedback based on Python interpreter errors requires running the code, which can result in undesirable side effects from unsuccessfully executed code. However, a shortcoming of this approach is that calling the LLM multiple times can be slow.
(3) \emph{Supervised fine-tuning:} For smaller models that struggle to adhere to the allowed subset of Python, we can use supervised fine-tuning to improve adherence.

We show that generating \toolname{} code via transpilation retains task performance comparable to that of Python, and we present several strategies for improving transpilation success. Specifically, we compare ordinary, unrestricted Python generation (``Python'') to Python generation restricted to the allowed subset, via simple prompt instructions (``Ours''), as well as using multi-turn feedback (``Multi-turn''). Finally, we consider fine-tuning a small model (i.e., \ModelSftStudent) using programs generated by a large model (i.e., \ModelSftTeacher). We report results for both the base small model (``Nano Base'') and the fine-tuned version (``Nano SFT''); for these results, we use \SftTrainSize{} examples for training and report results on a held-out subset of \SftHeldOutSize{} examples. Figure~\ref{tab:eval:generation} shows, for each approach and dataset, the fraction of tasks where the generated program executes without error (``Execution'') and where it produces the correct result (``Accuracy''). Execution errors can be due to the LLM failing to adhere to the allowed subset or due to runtime errors. Restricting to the Python subset (``Ours'') does not significantly change accuracy for either dataset, while multi-turn feedback actually improves accuracy for both datasets (``Multi-turn''). Finally, SFT improves accuracy for GQA; the AgentDojo dataset contains only \AdTaskCount{} tasks, which is not enough for SFT.

\section{Enhancing code actions by extending \toolname{}}

To demonstrate how features supporting code actions can easily be implemented in \toolname{}, we implement three such features: access control with batched requests, autoparallelized execution of external calls, and conformal prediction for uncertainty quantification.

\subsection{Batching access control queries}
\begin{algorithm}[t]
\caption{Pseudocode for the batched execution semantics.
}
\label{alg:main}
\scriptsize
\begin{minipage}[t]{0.49\linewidth}
\begin{algorithmic}
\vspace{0.5em}
\Function{RunBatchedApproval}{$P$}
\While{$P$ has dispatchable calls}
\State Identify dispatchable external calls $\{S\}$ in $P$
\State Query user to validate $\{S\}$
\State Terminate execution if rejected
\State Dispatch $\{S\}$ and add to a set $E$
\State $P\gets\textsc{RunInternal}(P,E)$
\EndWhile
\State \Return $P$
\EndFunction
\vspace{0.5em}
\end{algorithmic}
\end{minipage}
\begin{minipage}[t]{0.49\linewidth}
\begin{algorithmic}
\vspace{0.5em}
\Function{RunInternal}{$P,E$}
\While{\textbf{true}}
\If{exists $(y\gets f(x_1,...,x_k),B)\in E$ \textbf{and} $B\neq\varnothing$}
\State apply $R_{\text{ext}}$ to $P$ to substitute $B$ in for $y$
\ElsIf{exists rule $R\in\mathcal{R}_{\text{int}}$ applicable to $P$}
\State apply $R$ to $P$
\ElsIf{$E=\varnothing$}
\State \textbf{break}
\EndIf
\EndWhile
\State \Return $P$
\vspace{0.5em}
\EndFunction
\end{algorithmic}
\end{minipage}
\end{algorithm}

We consider a standard security model based on access control~\citep{accesscontrol,rbac}, where the user must approve the execution of effects. Because effects are isolated in external calls, we only need to ensure that external calls are consistent with the user's desired security policy. For instance, a smartphone user might grant an app access to resources such as their location and the ability to send emails, in which case the app would be allowed to access only these resources.

Because of the large space of actions and the uncertain risk tolerance of users, many systems, such as the coding agent Claude Code~\citep{claudecode}, dynamically ask users to approve each external call. A key challenge with dynamic access control is that queries can block progress while waiting for the user to approve requests, preventing agents from working autonomously. Thus, we implement a \emph{batched-approval semantics} that collects as many external calls as possible and queries the user to confirm them simultaneously (Algorithm~\ref{alg:main}). If the user rejects them, execution terminates; otherwise, the external calls are dispatched and execution proceeds. \textsc{RunInternal} performs as many rewrites of the current program $P$ as possible (including both applying internal rules and handling previously dispatched external calls). It returns once $P$ cannot be rewritten any further, at which point the main routine \textsc{RunBatchedApproval} queries the user to validate all the dispatchable external calls in $P$, and then dispatches all of these calls. This loop continues until $P$ is terminal. For example, in Figure~\ref{fig:main}, \toolname{} would ask the user for permission to make the external call \pyinline{image_patch.find("drink")} in $P_1$, but would then be able to batch the permission requests for \pyinline{patch1.simple_query(...)} and \pyinline{patch2.simple_query(...)} in $P_3$.

\subsection{Autoparallelizing external calls}
Many external tool calls (e.g., recursive calls to sub-agents) execute very slowly; thus, to reduce latency, they would ideally be executed in parallel where possible. The design of \toolname{} is naturally suited to this: because programs are interpreted using rewrite rules, a statement can be ``executed'' as soon as the relevant program variables are substituted with constants,
allowing statements to run out of order and in parallel. The example in Figure~\ref{fig:main} is actually executing with \toolname{}'s \emph{autoparallelization semantics}: in program $P_3$, the statement \pyinline{patch2.simple_query(...)} can be evaluated even though previous statements have not yet been evaluated, since all of its arguments (the image patch \pyinline{patch2} and the string \pyinline{"Does this have alcohol?"}) are constants. As a consequence, this external call can be dispatched in parallel with \pyinline{patch1.simple_query(...)}, which significantly reduces execution time compared to sequential Python.

\subsection{Conformal prediction}
\begin{figure}[t]
\centering
\includegraphics[width=0.35\textwidth]{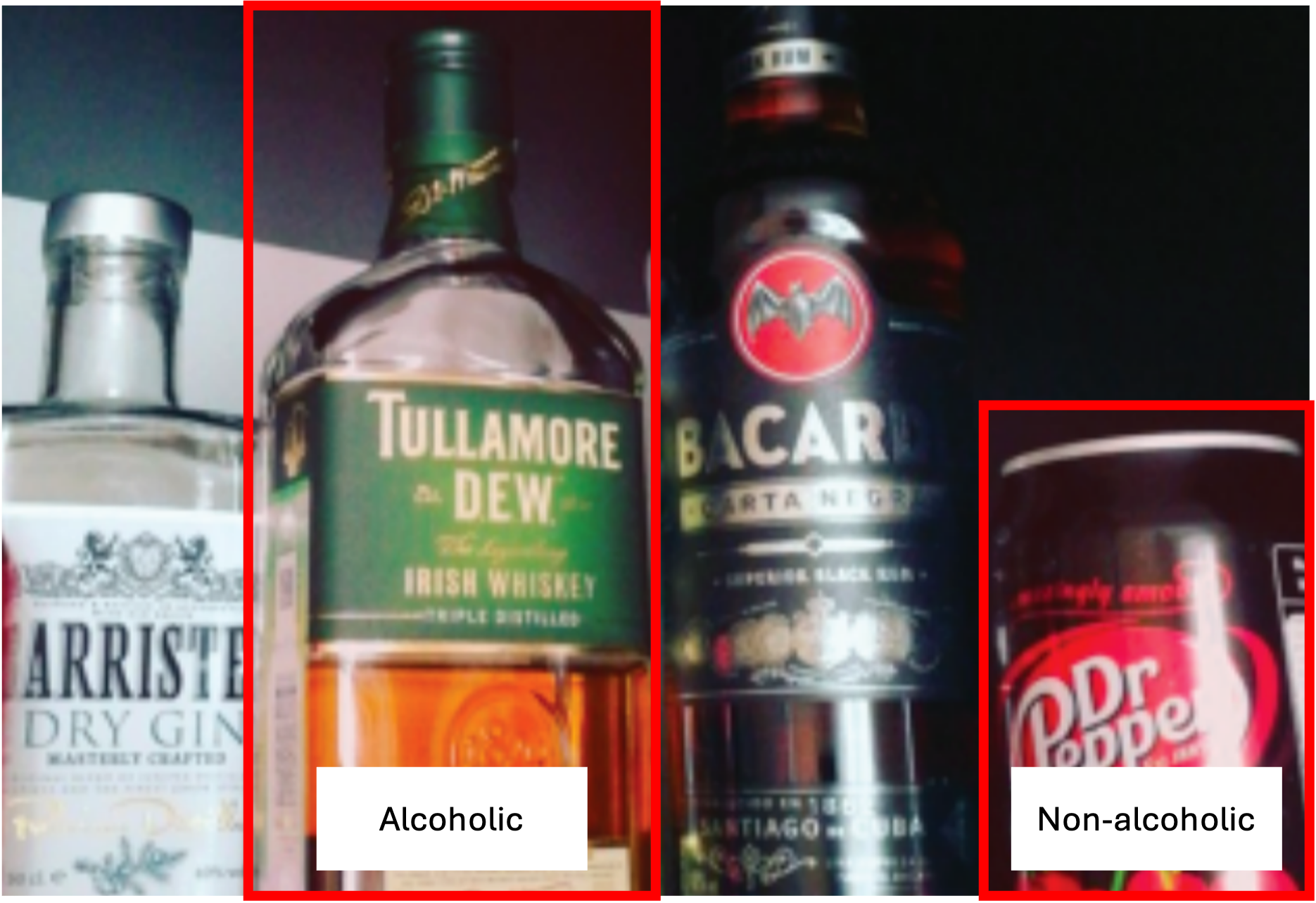}
\quad
\includegraphics[width=0.35\textwidth]{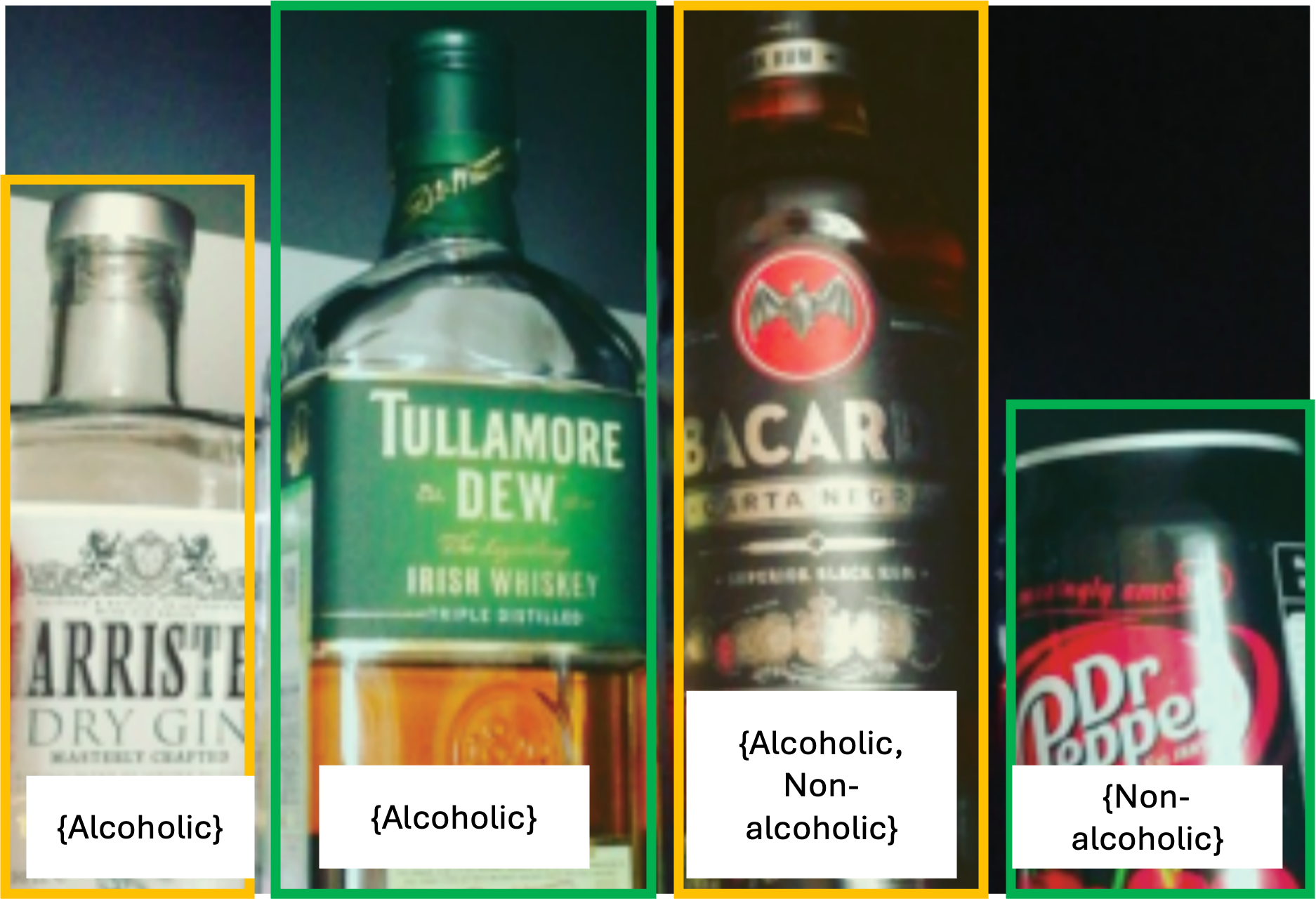}
\caption{Illustration of how the \pyinline{image_patch.find("drink")} call from Figure~\ref{fig:example} can be adapted from a standard object detector (left) to a conformal one (right). For the latter, the green boxes are identified as definitely being in the image, whereas the yellow boxes may or may not be in the image.}
\label{fig:example-conformal}
\end{figure}

We implement \emph{conformal semantics}~\citep{ramalingam2024} in \toolname{} for uncertainty quantification.
When \toolname{} makes external calls to other machine learning models, such as sub-agents, we may want to quantify the uncertainty of these models, and then keep track of how this uncertainty propagates through the program.
Specifically, program variables are bound to sets of values instead of individual values.

The key challenge is modifying the program execution to handle sets of values. For example, if a Boolean variable $x$ is bound to the set of values $x\mapsto\{\pyinlinem{True},\pyinlinem{False}\}$, and there is a conditional statement $\pyinlinem{if}\ x\ \pyinlinem{then}\ p_{\text{true}}\ \pyinlinem{else}\ p_{\text{false}}$ (i.e., that branches on $x$), then we must execute both branches $p_{\text{true}}$ and $p_{\text{false}}$ of the conditional; then, for each variable $y$ defined in these branches, we take the union of the values $v_{\text{true}}$ bound to $y$ in $p_{\text{true}}$ and $v_{\text{false}}$ bound to $y$ in $p_{\text{false}}$, i.e., $y\mapsto v_{\text{true}}\cup v_{\text{false}}$. \toolname{} includes a modified set of \emph{conformal} rewrite rules that handle variables bound to sets of values in this way (see Appendix~\ref{sec:fulllang-conformal}).

Because external functions are opaque to \toolname{}, abstract versions of them must be provided. In the case of calls to neural models, such as \pyinline{find}, the abstract version is provided by applying a conformal technique, e.g., returning the set of labels whose probability is above some threshold. For example, the object detector shown on the left in Figure~\ref{fig:example-conformal} misses two objects (though in this case, it does not affect the final answer in Figure~\ref{fig:main}); the output of the conformal detector is shown on the right. In this case, the external call \pyinline{image_patch.find("drink")} indicates whether each detection is definitely (green) or possibly (yellow) in the image; it represents the set of lists of patches
{\scriptsize
\begin{align*}
\{[\pyinlinem{patch1},\pyinlinem{patch2},\pyinlinem{patch3},\pyinlinem{patch4}], [\pyinlinem{patch2},\pyinlinem{patch3},\pyinlinem{patch4}], [\pyinlinem{patch1},\pyinlinem{patch2},\pyinlinem{patch4}],[\pyinlinem{patch2},\pyinlinem{patch4}]\},
\end{align*}
}%
where the patches are ordered from left to right. Similarly, for each patch, the external call \pyinline{patch.simple_query("Does this have alcohol?")} returns a prediction set that is a subset of
$\{\pyinlinem{"yes"},\pyinlinem{"no"}\}$.
The result overapproximates the true output; in this case, the program output is $\{\pyinlinem{"yes"}\}$, i.e., there is definitely an alcoholic drink in the image.

The conformal guarantee says that, for some target fraction of the test dataset (``coverage''), the ground truth label will be contained in the predicted set of labels. While this can be trivially obtained by outputting the set of all labels, the sets should be kept as small as possible while satisfying the target coverage.
To satisfy the desired coverage guarantee, we use a standard conformal prediction strategy. First, we optimize the thresholds for each individual model on an optimization set~\citep{traq}. Then, using a held-out calibration set, we jointly rescale these thresholds using a single scaling parameter $\tau\in\mathbb{R}$ chosen using conformal prediction to satisfy a desired coverage guarantee~\citep{learnthentest,zhang2025}. $\tau$ can also determine the number of programs to generate, in order to handle uncertainty at the program level~\citep{quachconformal}.

\begin{figure*}[t]
\hfill
\begin{subfigure}[b]{0.49\textwidth}
\centering
\includegraphics[width=\textwidth]{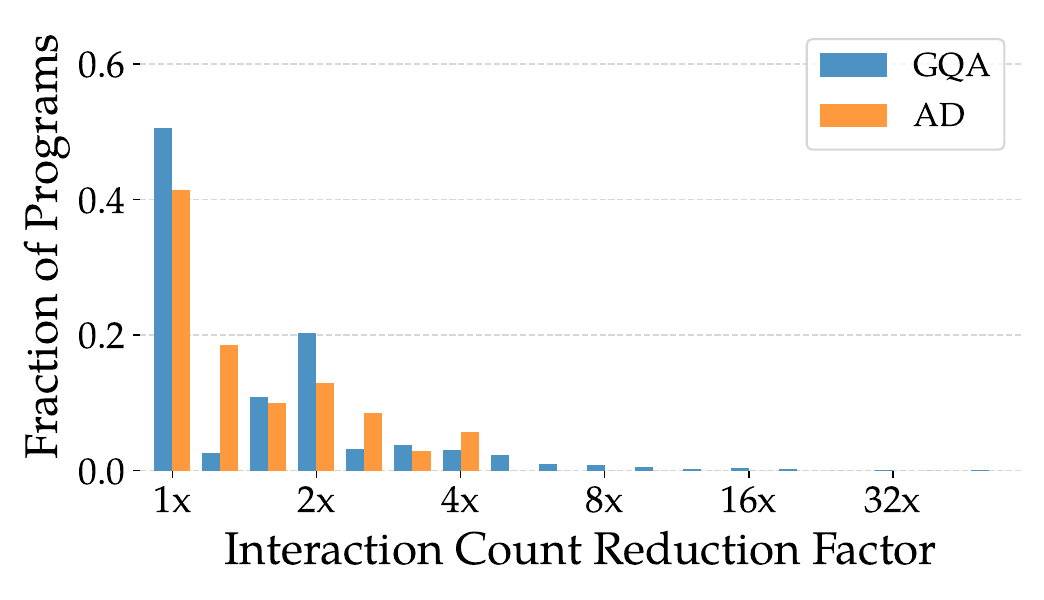}
\caption{}
\label{fig:results:security}
\end{subfigure}
\begin{subfigure}[b]{0.49\textwidth}
\centering
\includegraphics[width=\textwidth]{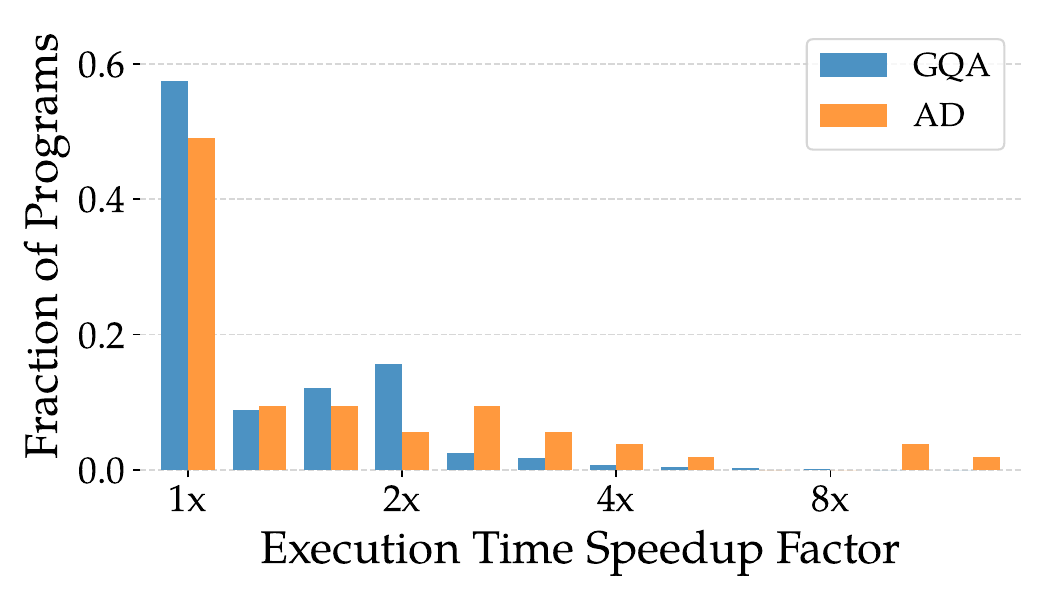}
\caption{}
\label{fig:results:performance}
\end{subfigure}
\hfill
\vspace{1em}\\\strut
\hfill
\begin{subfigure}[b]{0.9\textwidth}
\centering
\includegraphics[width=\textwidth]{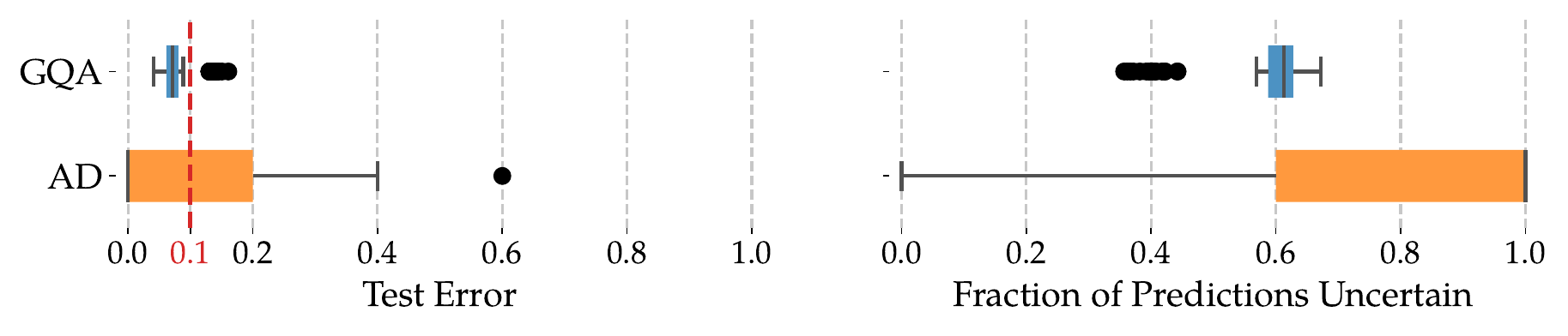}
\caption{}
\label{fig:results:reliability}
\end{subfigure}
\hfill\,
\caption{Distribution across tasks of interaction count reduction factors (Python count/\toolname{} count) from approval batching (a). Distribution across tasks of execution time reduction factor (Python time/\toolname{} time) from autoparallelization (b). Distributions across \ConfNumSplits{} validation/test splits of test error and uncertain prediction fraction for conformal semantics, satisfying a target \ConfTargetError{} error rate (c). The target error rate is shown as a dashed line.
}
\label{fig:results}
\end{figure*}

\section{Evaluation}
\label{sec:eval}

We show that \toolname{} significantly reduces the number of user interactions for access control, significantly reduces execution time, and achieves the target coverage rate using conformal prediction. We evaluate on ViperGPT~\citep{surís2023vipergptvisualinferencepython}, a visual question answering system, and CaMeL~\citep{camel}, a secure AI assistant. Given a natural language query about an image, ViperGPT first uses an LLM to generate a Python program that would answer that query when provided with an image. The Python program itself has access to various neural modules, including an object detector, a vision-language model, and an LLM. We apply the ViperGPT approach to \GqaTaskCount{} tasks randomly sampled from GQA~\citep{hudson2019gqa}, a dataset of questions about various day-to-day images. The CaMeL AI assistant performs tasks for users by using an LLM to generate Python code that itself may invoke ``quarantined'' LLMs to process data without risk of prompt injections. We apply the CaMeL approach to all four suites of the AgentDojo~\citep{debenedetti2024agentdojo} benchmark. Unless otherwise stated, all of our results use programs generated by \ModelCodegen{} at the default (``\ModelCodegenEffort{}'') reasoning level. We only include programs in our analysis that successfully compile and execute in \toolname{} (i.e., no compilation or execution errors), but we include programs regardless of whether their output matches the ground truth label.

\begin{figure*}[t]
\hfill
\begin{subfigure}[b]{0.49\textwidth}
\centering
\includegraphics[width=\textwidth]{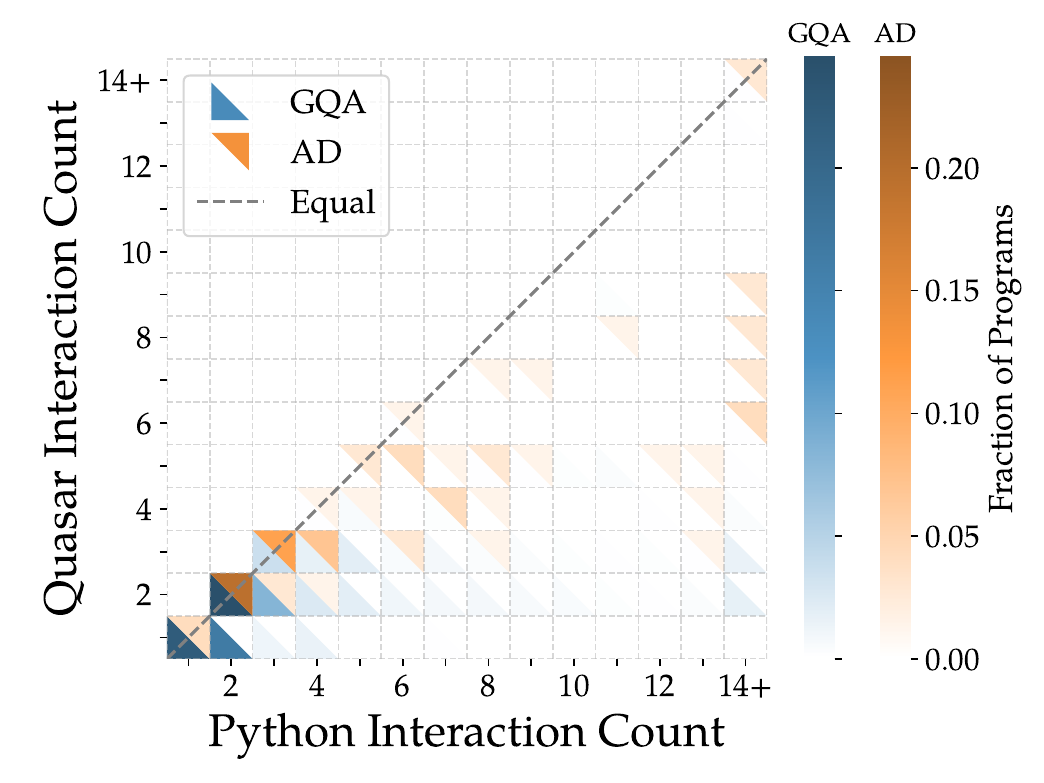}
\caption{}
\label{fig:eval:security}
\end{subfigure}
\begin{subfigure}[b]{0.49\textwidth}
\centering
\includegraphics[width=\textwidth]{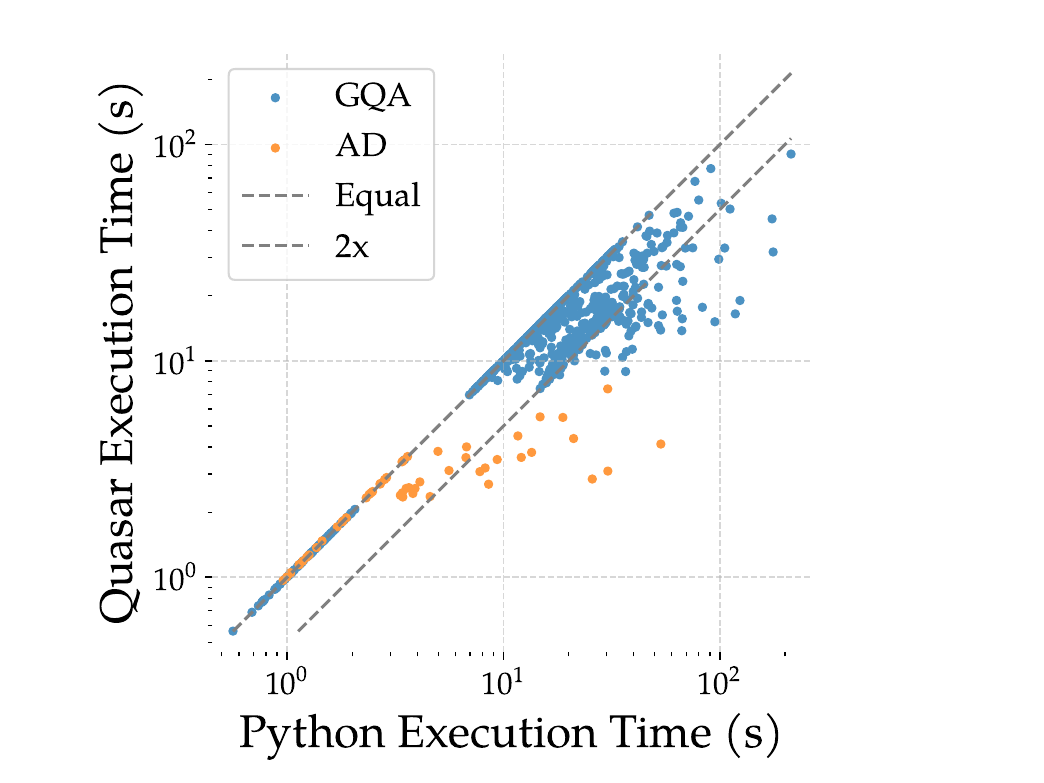}
\caption{}
\label{fig:eval:performance}
\end{subfigure}
\hfill\,
\caption{Python vs.\ \toolname{} user interactions required across tasks (a). Python vs.\ \toolname{} execution time across tasks (b).}
\label{fig:eval}
\end{figure*}

\textbf{Access control.}
We evaluate the access control improvements in terms of the reduction in the number of user interactions required to approve all external calls, comparing the original Python program to its transpiled \toolname{} counterpart. The transpiler we use guarantees that the Python source program and \toolname{} target program make exactly the same external calls. The distribution of improvements is shown in Figure~\ref{fig:results:security} and the interaction counts are shown in Figure~\ref{fig:eval:security}. On GQA, \toolname{} reduces the number of user interactions by \GqaRoundsOverall{} (mean $\pm$ stddev). The large variance is because only \GqaRoundsFracImprovable{} of tasks allow approvals to be batched; among those, the number of interactions is reduced by \GqaRoundsImprovable{}. On AgentDojo, the overall reduction is \AdRoundsOverall, with \AdRoundsFracImprovable{} of tasks improvable and a reduction of \AdRoundsImprovable{} on those. These results demonstrate that \toolname{} can substantially reduce the frequency with which execution must block pending user approval.

\textbf{Autoparallelization.}
We again consider pairs of Python programs and their transpiled \toolname{} counterparts. To reduce variance in the results, we control for the time that each external call takes to execute, recording every external call that a program makes along with its result and execution time. Then, we replay this recording on both the Python and \toolname{} versions of the program and record the total execution time of each. We exclude from the analysis programs that have negligible Python execution time ($\leq \SpeedupRuntimeFloor{}$). The distribution of improvements is shown in Figure~\ref{fig:results:performance} and the execution times are shown in Figure~\ref{fig:eval:performance}. On the GQA dataset, \toolname{} reduces running time by \GqaSpeedupOverall{}. The large variance is because only \GqaSpeedupFracImprovable{} of tasks are parallelizable; among those, the running time is cut by \GqaSpeedupImprovable. On AgentDojo, the overall speedup is \AdSpeedupOverall, with \AdSpeedupFracImprovable{} of tasks improvable and a speedup of \AdSpeedupImprovable{} on those. These results demonstrate that \toolname{} can substantially reduce running time compared to Python execution.

\textbf{Conformal semantics.}
We show that our conformal semantics can achieve a target error rate of \ConfTargetError{} on a test set. For each benchmark, we divide the set of programs (one per task) into \ConfNumSplits{} random validation/test splits. For each split, we choose the largest threshold (and thus the smallest prediction sets) for which the validation error is below \ConfTargetError, and then compute the test error at that threshold. The distribution of these test errors is shown in Figure~\ref{fig:results:reliability}, with mean error rate \GqaConfError{} for GQA and \AdConfError{} for AgentDojo. Because the domain of labels varies across tasks (e.g., yes/no, color, object, etc.), instead of measuring the size of prediction sets we measure certainty---i.e., the model is certain if the prediction set has size 1, and uncertain otherwise. We consider the fraction of tasks on which the model is uncertain. The distribution of such uncertainty rates is shown in Figure~\ref{fig:results:reliability}, with mean \GqaConfUncertain{} in GQA and \AdConfUncertain{} for AgentDojo. The high uncertainty rate for AgentDojo reflects the stringent certainty condition and the challenging nature of the dataset. These results demonstrate that \toolname{} can achieve the desired coverage guarantee using conformal prediction.

\textbf{Scaling to complex benchmarks.}
Finally, we perform an additional experiment to assess whether \toolname{} can scale to more complex reasoning tasks. Specifically, we apply it to the BrowseComp-Plus~\citep{chen2025browsecompplus} deep research dataset, which consists of question answering tasks where each task is accompanied by a large number ($\approx 100$) of documents. The answer to each question requires integrating information across multiple documents; however, most documents are unrelated to the question and must be ignored. We apply the Recursive Language Models (RLM)~\citep{zhang2025recursivelanguagemodels} approach, adapted to use \toolname{} instead of Python. In RLM, rather than passing the documents directly into the LLM context, which may exceed the context window or lead to ``context rot'' (where model performance generally degrades as context size increases), the LLM is given access to a Python interpreter in which the documents exist as variables. Thus, it can manipulate documents programmatically and pass some subset of them to another LLM for processing, without needing to pollute its own context. 

Due to the large number of model calls required for each task, we evaluate on a randomly selected subset of \BcpSubsetSize{}
BrowseComp-Plus tasks, using \BcpModel{} for the inner model calls. We see large improvements on this dataset: there is a \BcpRoundsOverall{} reduction in the number of user approvals required and a \BcpSpeedupOverall{} speedup. Evaluating conformal prediction on this benchmark is complicated by the unstructured nature of the intermediate values and outputs, so we leave it to future work.

BrowseComp-Plus represents a realistic deep research benchmark, so these results demonstrate that \toolname{} straightforwardly scales to complex tasks, with only the conformal prediction component requiring a domain-specific implementation. Indeed, in terms of latency and security improvements, we find stronger results on these complex tasks than on GQA and AgentDojo; complex tasks require comparatively more external calls, creating more opportunities for parallelization and batching.

\section{Conclusion}
\label{sec:conclimit}

We have presented \toolname{}, a language for LLM code actions that distinguishes between internal code and external calls. Leveraging LLMs' proficiency with Python, we transpile from a subset of Python into \toolname{}.
Unlike Python, \toolname{} has semantics that can be extended to support a wide variety of features. We demonstrate extensions for access control approval batching, automatic parallelization, and conformal prediction.
Our experiments demonstrate that via these semantics, \toolname{} can improve security, latency, and uncertainty quantification on both GQA and AgentDojo, and can furthermore scale to significantly more complex tasks such as BrowseComp-Plus.

%% file: plots/eval_generation.tex
\begin{tabular}{lrrrr}
\toprule
\textbf{Approach} & \multicolumn{2}{c}{\textbf{Execution}} & \multicolumn{2}{c}{\textbf{Accuracy}} \\
\cmidrule{2-3}\cmidrule{4-5} & \textbf{GQA} & \textbf{AD} & \textbf{GQA} & \textbf{AD} \\
\midrule
Python & 99.6 & 76.3 & 71.8 & 64.5 \\
Ours & 97.9 & 82.8 & 71.4 & 63.4 \\
\midrule
Multi-turn & 99.5 & 89.2 & 72.3 & 67.7 \\
\midrule
Nano Base & 92 & \multicolumn{1}{c}{--} & 65 & \multicolumn{1}{c}{--} \\
Nano SFT & 99 & \multicolumn{1}{c}{--} & 71 & \multicolumn{1}{c}{--} \\
\bottomrule
\end{tabular}

%% file: appendix-functionalization.tex
\clearpage
\section{Full \toolname{} language}
\label{sec:fulllang}
As described in Section~\ref{sec:lang}, \toolname{} executes programs by transforming them with rewrite rules until they reach a result. The syntax of programs is given in Figure~\ref{fig:fulllang:syntax}. A program consists of a sequence of statements, where each statement defines some variable ($x, y, \ldots$) to be the result of some operation ($op$). Variables are assumed to be defined exactly once (i.e., they are unique and do not shadow each other). An operation can be, in order: a primitive value $c$ (where $c$ ranges over Python values, such as \pyinline{True}, \pyinline{5}, or \pyinline{"foo"}); another variable $x$; a tuple of variables $x_i$; the result of calling an external function $f$ (from the set $\mathcal{F}_{\text{ext}}$) with argument $x$; the result of projecting out the $i$-th component from a tuple $x$; the result of folding over a list $w$ with initial accumulator $x$ and fold body $block$; an if expression on condition $x$ with then-case $block_1$ and else-case $block_2$; or the result of some pending external call $S$. A block is a program that may additionally have some parameter $x$ (so that the body of a fold can take the previous accumulator and the current list item as arguments). A value is either a Python object or a (possibly nested) tuple of Python objects---it does not directly occur in programs, but is used in the semantics.

The interpreter state at any time is simply a program $P$ and a set $E$ of dispatched external calls. The semantics consists of rewrite rules $R \in \mathcal{R}$, which transform one execution state into another, written $P, E \rightarrow_{R} P', E'$. Many rules do not affect $E$, and so are simply written as $P \rightarrow_{R} P'$.
The rewrite rules are given in Figure~\ref{fig:fulllang:rules}. The rule ``alias'' removes a statement $y \gets x$, replacing it with nothing, but renaming all occurrences of $y$ in the program to $x$; ``proj'' replaces a projection operator, if the variable $x$ is known to be a tuple $(x_1, \ldots, x_n)$, with the $i$-th element; for if statements, when the condition $x$ is the primitive \pyinline{True} (``if-t''), then the statement is replaced by a copy of $block_1$ (copying ensures that variables are unique; since blocks in if statements do not require parameters, $w$ is bound to an empty tuple); if the condition is \pyinline{False} (``if-f''), then the same is done for $block_2$; ``fold'' applies $block$ to each element of the list $w$, with $x$ being the initial accumulator and $y$ being the final one, and $z_i$ being the $i$-th intermediate accumulator; ``disp'' replaces an external call to a function $f$ when the argument $x$ has a value $value$ (i.e., it is a primitive or a tuple of primitives, which $\operatorname{value}(T, x)$ computes) with a placeholder $S$, begins executing the function $f$, and updates the execution set $E$; ``ext'' applies when an external function has finished executing---and so the execution set $E$ contains a result in place of $\varnothing$---and replaces the placeholder with the result. In Section~\ref{sec:lang}, we simplified $S$ in the execution set to just be the external call statement itself, whereas here it is an identifier for the spawned task.

\begin{figure}[hp]
\small
\centering
\begin{align*}
P \Coloneqq&\ stmt_1; \ldots; stmt_n; \operatorname{return} x \\
stmt \Coloneqq&\ x \gets op \\
op \Coloneqq&\ \operatorname{prim}\ c \\
\mid&\ x \\
\mid&\ (x_1, \ldots, x_n) \\
\mid&\ f\ x \\
\mid&\ \operatorname{proj}\ i\ x \\
\mid&\ \operatorname{fold}\ w\ x\ block \\
\mid&\ \operatorname{if}\ x\ block_1\ block_2 \\
\mid&\ {?S} \\
block \Coloneqq&\ \{x \Rightarrow P\} \\
value \Coloneqq&\ c \mid (value_1, \ldots, value_n)
\end{align*}
\caption{The grammar defining programs in \toolname{}.}
\label{fig:fulllang:syntax}
\end{figure}

\begin{figure}[hp]
\small
\centering
\begin{subfigure}[b]{\textwidth}
\begin{mathpar}
\inferrule{
}{
T\mathopen{}\left\lbrack y \gets x\right\rbrack\mathclose{} \rightarrow \operatorname{rename}(T\mathopen{}\left\llbracket \varnothing\right\rrbracket\mathclose{}, y, x)
} \text{(alias)} \and
\inferrule{
(x \gets (x_1, \ldots, x_n)) \in T
}{
T\mathopen{}\left\lbrack y \gets \operatorname{proj}\ i\ x\right\rbrack\mathclose{} \rightarrow T\mathopen{}\left\llbracket y \gets x_i\right\rrbracket\mathclose{}
} \text{(proj)} \and
\inferrule{
(x \gets \operatorname{prim}\ \pyinlinem{True}) \in T \\
\{w \Rightarrow stmts; \operatorname{return} z\} = \operatorname{copy}(block_1)
}{
T\mathopen{}\left\lbrack y \gets \operatorname{if}\ x\ block_1\ block_2\right\rbrack\mathclose{} \rightarrow T\mathopen{}\left\llbracket w \gets (); stmts; y \gets z\right\rrbracket\mathclose{}
} \text{(if-t)} \and
\inferrule{
(x \gets \operatorname{prim}\ \pyinlinem{False}) \in T \\
\{w \Rightarrow stmts; \operatorname{return} z\} = \operatorname{copy}(block_2)
}{
T\mathopen{}\left\lbrack y \gets \operatorname{if}\ x\ block_1\ block_2\right\rbrack\mathclose{} \rightarrow T\mathopen{}\left\llbracket w \gets (); stmts; y \gets z\right\rrbracket\mathclose{}
} \text{(if-f)} \and
\inferrule{
(w \gets \operatorname{prim}\ [c_1, \ldots, c_n]) \in T \\
\forall i. \{y_i \Rightarrow stmts_i; \operatorname{return} z_i\} = \operatorname{copy}(block) \\
stmts'_i = (w_i \gets \operatorname{prim}\ c_i; y_i \gets (z_{i-1},w_i); stmts_i)
}{
T\mathopen{}\left\lbrack y \gets \operatorname{fold}\ w\ x\ block\right\rbrack\mathclose{} \rightarrow T\mathopen{}\left\llbracket z_0 \gets x; stmts_1'; \ldots; stmts_n'; y \gets z_n\right\rrbracket\mathclose{}
} \text{(fold)} \and
\inferrule{
\operatorname{value}(T, x) = value \\
S = \operatorname{spawn}(f, value)
}{
T\mathopen{}\left\lbrack y \gets f\ x\right\rbrack\mathclose{}, E \rightarrow T\mathopen{}\left\llbracket y \gets {?S}\right\rrbracket\mathclose{}, E \cup \{(S, \varnothing)\}
} \text{(disp)} \and
\inferrule{
term = (stmts; \operatorname{return} x)
}{
T\mathopen{}\left\lbrack y \gets {?S}\right\rbrack\mathclose{}, E \cup \{(S, term)\} \rightarrow T\mathopen{}\left\llbracket stmts; y \gets x\right\rrbracket\mathclose{}, E
} \text{(ext)}
\end{mathpar}
\caption{}
\end{subfigure}
\begin{subfigure}[b]{\textwidth}
\begin{mathpar}
\inferrule{
(x \gets \operatorname{prim}\ c) \in T
}{
\operatorname{value}(T, x) = c
} \and
\inferrule{
(x \gets (x_1, \ldots, x_n)) \in T \\
\forall i. \operatorname{value}(T, x_i) = value_i
}{
\operatorname{value}(T, x) = (value_1, \ldots, value_n)
}
\end{mathpar}
\caption{}
\end{subfigure}
\caption{The rewrite rules of the semantics of \toolname{} (a), and the formal definition of the $\operatorname{value}$ function used by the ``disp'' rule (b). $T\mathopen{}\left\lbrack stmt\right\rbrack\mathclose{}$ means a program $P$ with some statement $stmt$ in it; $T\mathopen{}\left\llbracket stmts\right\rrbracket\mathclose{}$ means that the statement $stmt$ was replaced by the list of statements $stmts$. For each rule, the $P, E \rightarrow P', E'$ below the line is an allowed rewrite, subject to all of the conditions above the line. If $E$ is not modified by a rule, we omit it for concision. $stmt \in T$ means that $stmt$ is in the list of statements of $T$.}
\label{fig:fulllang:rules}
\end{figure}

\clearpage
\section{Transpilation}
\label{sec:transpilation}

\toolname{} is functional, while Python is imperative. Thus, in \toolname{}, variables cannot be changed once they have been defined. Being functional makes supporting parallel, partial, and conformal execution possible, but the Python code generated by agents performs imperative updates to local variables. Handling updates in ``straight-line'' code (without control-flow structures like \pyinline{if} and \pyinline{for}) is straightforward. However, updates inside control-flow structures, which therefore may or may not happen, are more challenging. In the example from Figure~\ref{fig:main}, \pyinline{found} may be updated inside the loop.

We adapt an existing transpiler~\citep{poppy} to translate to \toolname{} an expressive subset of Python that includes function calls, local variable assignments, and the \pyinline{if} and \pyinline{for} control-flow constructs.
To support imperative control-flow structures in Python, we convert them into a functional form. \pyinline{if} statements in Python are transformed as shown in Figure~\ref{fig:transpilation-if}, where variables that might be updated by a statement-level conditional are instead returned from an expression-level conditional. A similar translation is done from imperative \pyinline{for} loops to functional fold operations: variables that might be updated by the \pyinline{for} loop are instead passed as the fold accumulator (in Python, fold is called \pyinline{reduce}).

\begin{figure}[hp]
\scriptsize
\centering
\hfill
\begin{subfigure}[c]{0.34\textwidth}
\begin{minted}{python3}
found = False
cond = drink_patch.simple_query("...")
if cond:
    found = True

\end{minted}
\caption{}
\end{subfigure} \hfill $\rightsquigarrow$ \hfill
\begin{subfigure}[c]{0.4\textwidth}
\begin{minted}{python3}
found0 = False
cond = drink_patch.simple_query("...")
def then_case():
    found1 = True
    return found1
def else_case():
    return found0
found2 = then_case() if cond else else_case()
\end{minted}
\caption{}
\end{subfigure}
\hfill\,
\caption{An illustration of the translation of ``if'' statements, shown in Python syntax. Before, with imperative updates in ``if'' statements (a). After, with no imperative updates and a functional conditional operation (b).}
\label{fig:transpilation-if}
\end{figure}

\clearpage
\section{Conformal semantics for \toolname{}}
\label{sec:fulllang-conformal}

To support conformal evaluation, \toolname{} must be extended to handle sets of values. The syntax has three additional operations, as shown in Figure~\ref{fig:fulllang:conformal:syntax}. In order: abstract primitives represent the set of Python values $c_i$; abstract lists represent a list where some of the elements may be uncertain, in that if $b_i$ is \pyinline{False}, the element $c_i$ may or may not be in the list, whereas if $b_i$ is \pyinline{True}, then $c_i$ is definitely in the list; and join operations combine multiple computations into a set. Join is distinct from an abstract set, since in the latter the values must be known, whereas in the former they may not yet be computed.

The semantics also contains additional rules to support these new operations, as shown in Figure~\ref{fig:fulllang:conformal:rules}. The rule ``join-join'' applies when $x$ is the join of variables, and one of them, $y$, is itself a join, in which case they can be flattened to a single join; ``join-tuple'' applies when $x$ is the join of $n$ tuples of identical length $m$, in which case it becomes the tuple of joins of the respective components; ``join-prim'' applies when $x$ is the join of $n$ primitives $c_i$, in which case it becomes an abstract set of those values; ``if-tf'' applies when the condition of an if statement is the abstract set of both \pyinline{True} and \pyinline{False}, in which case both branches are taken, resulting in $z_1$ and $z_2$, which are joined to produce $y$; ``fold-abs'' applies when folding over an abstract list, in which case a copy of $block$ is made for each element of the list; however, if a list element is uncertain ($b_i = \pyinlinem{False}$), then the resulting accumulator $z_i$ is joined with $z_{i-1}$ to capture both the case where $c_i$ is in the list and the case where it is not.

\begin{figure}[hp]
\small
\centering
\begin{align*}
op \Coloneqq&\ \ldots \\
\mid&\ \operatorname{absprim}\ \{c_1, \ldots, c_n\} \\
\mid&\ \operatorname{abslist}\ [(c_1, b_1), \ldots, (c_n, b_n)] \\
\mid&\ \operatorname{join}\ \{x_1, \ldots, x_n\} \\
absvalue \Coloneqq&\ \{c_1, \ldots, c_n\} \mid (absvalue_1, \ldots, absvalue_n)
\end{align*}
\caption{The additional operations in the grammar of \toolname{} to support conformal evaluation.}
\label{fig:fulllang:conformal:syntax}
\end{figure}

\begin{figure}[hp]
  \small
  \centering
  \begin{mathpar}
    \inferrule{
      (y \gets \operatorname{join}\ \{y_1, \ldots, y_m\}) \in T
    }{
      T\mathopen{}\left\lbrack x \gets \operatorname{join}\ \{x_1, \ldots, x_n, y\}\right\rbrack\mathclose{} \rightarrow T\mathopen{}\left\llbracket x \gets \operatorname{join}\ \{x_1, \ldots, x_n, y_1, \ldots, y_m\}\right\rrbracket\mathclose{}
    } \text{(join-join)} \and
    \inferrule{
      \forall i. (x_i \gets (w_{i,1}, \ldots, w_{i,m})) \in T \\
      stmts_j = (y_j \gets \operatorname{join}\ \{w_{1, j}, \ldots, w_{n, j}\})
    }{
      T\mathopen{}\left\lbrack x \gets \operatorname{join}\ \{x_1, \ldots, x_n\}\right\rbrack\mathclose{} \rightarrow T\mathopen{}\left\llbracket stmts_1; \ldots; stmts_m; x \gets (y_1, \ldots, y_m)\right\rrbracket\mathclose{}
    } \text{(join-tuple)} \and
    \inferrule{
      \forall i. (x_i \gets \operatorname{prim}\ c_i) \in T
    }{
      T\mathopen{}\left\lbrack x \gets \operatorname{join}\ \{x_1, \ldots, x_n\}\right\rbrack\mathclose{} \rightarrow T\mathopen{}\left\llbracket x \gets \operatorname{absprim}\ \{c_1, \ldots, c_n\}\right\rrbracket\mathclose{}
    } \text{(join-prim)} \and
    \inferrule{
      (x \gets \operatorname{absprim}\ \{\pyinlinem{True}, \pyinlinem{False}\}) \in T \\
      \{w_1 \Rightarrow stmts_1; \operatorname{return} z_1\} = \operatorname{copy}(block_1) \\
      \{w_2 \Rightarrow stmts_2; \operatorname{return} z_2\} = \operatorname{copy}(block_2)
    }{
      T\mathopen{}\left\lbrack y \gets \operatorname{if}\ x\ block_1\ block_2\right\rbrack\mathclose{} \rightarrow T\mathopen{}\left\llbracket w_1 \gets (); stmts_1; w_2 \gets (); stmts_2; y \gets \operatorname{join}\ \{z_1, z_2\}\right\rrbracket\mathclose{}
    } \text{(if-tf)} \and
    \inferrule{
      (w \gets \operatorname{abslist}\ [(c_1, b_1), \ldots, (c_n, b_n)]) \in T \\
      \forall i. \{y_i \Rightarrow stmts_i; \operatorname{return} z_i\} = \operatorname{copy}(block) \\\\
      stmts'_i = (w_i \gets \operatorname{prim}\ c_i; y_i \gets (z_{i-1},w_i); stmts_i) \\\\
      stmts''_i = \operatorname{if}\ b_i\ \operatorname{then}\ stmts'_i\ \operatorname{else}\ (stmts'_i; z_i \gets \operatorname{join}\ \{z_i, z_{i-1}\})
    }{
      T\mathopen{}\left\lbrack y \gets \operatorname{fold}\ w\ x\ block\right\rbrack\mathclose{} \rightarrow T\mathopen{}\left\llbracket z_0 \gets x; stmts_1''; \ldots; stmts_n''; y \gets z_n\right\rrbracket\mathclose{}
    } \text{(fold-abs)}
  \end{mathpar}
  \caption{The additional rewrite rules in the semantics of \toolname{} to support conformal evaluation.}
  \label{fig:fulllang:conformal:rules}
\end{figure}

%% file: main.bbl
\begin{thebibliography}{39}
\providecommand{\natexlab}[1]{#1}
\providecommand{\url}[1]{\texttt{#1}}
\expandafter\ifx\csname urlstyle\endcsname\relax
  \providecommand{\doi}[1]{doi: #1}\else
  \providecommand{\doi}{doi: \begingroup \urlstyle{rm}\Url}\fi

\bibitem[Andriushchenko et~al.(2025)Andriushchenko, Souly, Dziemian, Duenas, Lin, Wang, Hendrycks, Zou, Kolter, Fredrikson, Gal, and Davies]{andriushchenko2025agentharmbenchmarkmeasuringharmfulness}
Maksym Andriushchenko, Alexandra Souly, Mateusz Dziemian, Derek Duenas, Maxwell Lin, Justin Wang, Dan Hendrycks, Andy Zou, Zico Kolter, Matt Fredrikson, Yarin Gal, and Xander Davies.
\newblock {AgentHarm}: A benchmark for measuring harmfulness of {LLM} agents.
\newblock In \emph{International Conference on Learning Representations}, pp.\  79185--79220, 2025.

\bibitem[Angelopoulos et~al.(2025)Angelopoulos, Bates, Candès, Jordan, and Lei]{learnthentest}
Anastasios~N. Angelopoulos, Stephen Bates, Emmanuel~J. Candès, Michael~I. Jordan, and Lihua Lei.
\newblock Learn then test: Calibrating predictive algorithms to achieve risk control.
\newblock \emph{The Annals of Applied Statistics}, 19\penalty0 (2), 2025.
\newblock \doi{10.1214/24-AOAS1998}.

\bibitem[Anthropic(2024)]{computeruse}
Anthropic.
\newblock Introducing computer use, a new {Claude 3.5 Sonnet}, and {Claude 3.5 Haiku}, 2024.
\newblock URL \url{https://www.anthropic.com/news/3-5-models-and-computer-use}.

\bibitem[{Anthropic}(2026)]{claudecode}
{Anthropic}.
\newblock {Claude Code}, 2026.
\newblock URL \url{https://claude.com/product/claude-code}.
\newblock AI-powered coding assistant for developers.

\bibitem[Beautement et~al.(2008)Beautement, Sasse, and Wonham]{beautement2008compliance}
Adam Beautement, M.~Angela Sasse, and Mike Wonham.
\newblock The compliance budget: Managing security behaviour in organisations.
\newblock In \emph{Proceedings of the 2008 New Security Paradigms Workshop (NSPW)}, pp.\  47--58, 2008.
\newblock \doi{10.1145/1595676.1595684}.
\newblock URL \url{https://www.nspw.org/papers/2008/nspw2008-beautement.pdf}.

\bibitem[Cassano et~al.(2023)Cassano, Gouwar, Nguyen, Nguyen, Phipps-Costin, Pinckney, Yee, Zi, Anderson, Feldman, Guha, Greenberg, and Jangda]{cassano2023multiple}
Federico Cassano, John Gouwar, Daniel Nguyen, Sydney Nguyen, Luna Phipps-Costin, Donald Pinckney, Ming-Ho Yee, Yangtian Zi, Carolyn~Jane Anderson, Molly~Q Feldman, Arjun Guha, Michael Greenberg, and Abhinav Jangda.
\newblock {MultiPL-E}: A scalable and polyglot approach to benchmarking neural code generation.
\newblock \emph{IEEE Transactions on Software Engineering}, 49\penalty0 (7):\penalty0 3675--3691, 2023.
\newblock \doi{10.1109/TSE.2023.3267446}.
\newblock URL \url{https://arxiv.org/abs/2208.08227}.

\bibitem[Chen et~al.(2025)Chen, Ma, Zhuang, Nie, Zou, Liu, Green, Patel, Meng, Su, Sharifymoghaddam, Li, Hong, Shi, Liu, Thakur, Zhang, Gao, Chen, and Lin]{chen2025browsecompplus}
Zijian Chen, Xueguang Ma, Shengyao Zhuang, Ping Nie, Kai Zou, Andrew Liu, Joshua Green, Kshama Patel, Ruoxi Meng, Mingyi Su, Sahel Sharifymoghaddam, Yanxi Li, Haoran Hong, Xinyu Shi, Xuye Liu, Nandan Thakur, Crystina Zhang, Luyu Gao, Wenhu Chen, and Jimmy Lin.
\newblock {BrowseComp-Plus}: A more fair and transparent evaluation benchmark of deep-research agent, 2025.
\newblock URL \url{https://arxiv.org/abs/2508.06600}.

\bibitem[Debenedetti et~al.(2024)Debenedetti, Zhang, Balunovic, Beurer-Kellner, Fischer, and Tram{\`e}r]{debenedetti2024agentdojo}
Edoardo Debenedetti, Jie Zhang, Mislav Balunovic, Luca Beurer-Kellner, Marc Fischer, and Florian Tram{\`e}r.
\newblock {AgentDojo}: A dynamic environment to evaluate prompt injection attacks and defenses for {LLM} agents.
\newblock In \emph{Advances in Neural Information Processing Systems (Datasets and Benchmarks Track)}, volume~37, pp.\  82895--82920, 2024.
\newblock \doi{10.52202/079017-2636}.

\bibitem[Debenedetti et~al.(2026)Debenedetti, Shumailov, Fan, Hayes, Carlini, Fabian, Kern, Shi, Terzis, and Tramèr]{camel}
Edoardo Debenedetti, Ilia Shumailov, Tianqi Fan, Jamie Hayes, Nicholas Carlini, Daniel Fabian, Christoph Kern, Chongyang Shi, Andreas Terzis, and Florian Tramèr.
\newblock Defeating prompt injections by design.
\newblock In \emph{4th IEEE Conference on Secure and Trustworthy Machine Learning (SaTML)}, 2026.

\bibitem[He et~al.(2026)He, Zhu, Ye, Liu, Zhou, and Yu]{he2026emergedsecurityprivacyllm}
Feng He, Tianqing Zhu, Dayong Ye, Bo~Liu, Wanlei Zhou, and Philip~S. Yu.
\newblock The emerged security and privacy of {LLM} agent: A survey with case studies.
\newblock \emph{ACM Computing Surveys}, 58\penalty0 (6):\penalty0 162:1--162:36, 2026.
\newblock \doi{10.1145/3773080}.

\bibitem[Hudson \& Manning(2019)Hudson and Manning]{hudson2019gqa}
Drew~A Hudson and Christopher~D Manning.
\newblock {GQA}: A new dataset for real-world visual reasoning and compositional question answering.
\newblock In \emph{Proceedings of the IEEE/CVF Conference on Computer Vision and Pattern Recognition (CVPR)}, pp.\  6700--6709, 2019.
\newblock \doi{10.1109/CVPR.2019.00686}.

\bibitem[Hughes(2026)]{automode}
John Hughes.
\newblock How we built {Claude Code} auto mode: a safer way to skip permissions.
\newblock \url{https://www.anthropic.com/engineering/claude-code-auto-mode}, Mar 2026.
\newblock Accessed: 2026-08-08.

\bibitem[Lewis et~al.(2020)Lewis, Perez, Piktus, Petroni, Karpukhin, Goyal, Küttler, Lewis, tau Yih, Rocktäschel, Riedel, and Kiela]{lewis2020retrievalaugmentedgenerationknowledgeintensivenlp}
Patrick Lewis, Ethan Perez, Aleksandra Piktus, Fabio Petroni, Vladimir Karpukhin, Naman Goyal, Heinrich Küttler, Mike Lewis, Wen tau Yih, Tim Rocktäschel, Sebastian Riedel, and Douwe Kiela.
\newblock Retrieval-augmented generation for knowledge-intensive {NLP} tasks.
\newblock In \emph{Advances in Neural Information Processing Systems}, volume~33, pp.\  9459--9474, 2020.

\bibitem[Li et~al.(2024{\natexlab{a}})Li, Kan, Callot, Bhasker, Rashid, and Esler]{li2024redoexecutionfreeruntimeerror}
Shou Li, Andrey Kan, Laurent Callot, Bhavana Bhasker, Muhammad~Shihab Rashid, and Timothy~B Esler.
\newblock {REDO}: Execution-free runtime error detection for {COding} agents, 2024{\natexlab{a}}.
\newblock URL \url{https://arxiv.org/abs/2410.09117}.

\bibitem[Li et~al.(2024{\natexlab{b}})Li, Park, Lee, and Bastani]{traq}
Shuo Li, Sangdon Park, Insup Lee, and Osbert Bastani.
\newblock {TRAQ}: Trustworthy retrieval augmented question answering via conformal prediction.
\newblock In \emph{Proceedings of the 2024 Conference of the North American Chapter of the Association for Computational Linguistics: Human Language Technologies (Volume 1: Long Papers)}, pp.\  3799--3821, 2024{\natexlab{b}}.
\newblock \doi{10.18653/v1/2024.naacl-long.210}.

\bibitem[Liu et~al.(2024{\natexlab{a}})Liu, Pan, Li, and Chen]{liu2024uncertaintyestimationquantificationllms}
Linyu Liu, Yu~Pan, Xiaocheng Li, and Guanting Chen.
\newblock Uncertainty estimation and quantification for {LLMs}: A simple supervised approach, 2024{\natexlab{a}}.
\newblock URL \url{https://arxiv.org/abs/2404.15993}.

\bibitem[Liu et~al.(2024{\natexlab{b}})Liu, Chen, Tian, Zou, Chen, and Cui]{liu2024llmconversationalagentmemory}
Na~Liu, Liangyu Chen, Xiaoyu Tian, Wei Zou, Kaijiang Chen, and Ming Cui.
\newblock From {LLM} to conversational agent: A memory enhanced architecture with fine-tuning of large language models, 2024{\natexlab{b}}.
\newblock URL \url{https://arxiv.org/abs/2401.02777}.

\bibitem[Maharana et~al.(2024)Maharana, Lee, Tulyakov, Bansal, Barbieri, and Fang]{maharana2024evaluatinglongtermconversationalmemory}
Adyasha Maharana, Dong-Ho Lee, Sergey Tulyakov, Mohit Bansal, Francesco Barbieri, and Yuwei Fang.
\newblock Evaluating very long-term conversational memory of {LLM} agents.
\newblock In \emph{Proceedings of the 62nd Annual Meeting of the Association for Computational Linguistics (Volume 1: Long Papers)}, pp.\  13851--13870, 2024.
\newblock \doi{10.18653/v1/2024.acl-long.747}.

\bibitem[Mell et~al.(2025)Mell, Kallas, Zdancewic, and Bastani]{opal}
Stephen Mell, Konstantinos Kallas, Steve Zdancewic, and Osbert Bastani.
\newblock Opportunistically parallel lambda calculus.
\newblock \emph{Proc. ACM Program. Lang.}, 9\penalty0 (OOPSLA2), October 2025.
\newblock \doi{10.1145/3763143}.

\bibitem[Mell et~al.(2026{\natexlab{a}})Mell, Mell, Kallas, Zdancewic, and Bastani]{poppy}
Stephen Mell, David Mell, Konstantinos Kallas, Steve Zdancewic, and Osbert Bastani.
\newblock {PopPy}: Opportunistically exploiting parallelism in {Python} compound {AI} applications, 2026{\natexlab{a}}.
\newblock URL \url{https://arxiv.org/abs/2605.18697}.

\bibitem[Mell et~al.(2026{\natexlab{b}})Mell, Zhang, and Mell]{artifact}
Stephen Mell, Botong Zhang, and David Mell.
\newblock Artifact for "{Quasar}: {A} {Programming} {Language} {Specialized} for {LLM} {Code} {Actions}", August 2026{\natexlab{b}}.
\newblock URL \url{https://doi.org/10.5281/zenodo.22086130}.

\bibitem[Orlanski et~al.(2023)Orlanski, Xiao, Garcia, Hui, Howland, Malmaud, Austin, Singh, and Catasta]{orlanski2023measuring}
Gabriel Orlanski, Kefan Xiao, Xavier Garcia, Jeffrey Hui, Joshua Howland, Jonathan Malmaud, Jacob Austin, Rishabh Singh, and Michele Catasta.
\newblock Measuring the impact of programming language distribution.
\newblock In \emph{Proceedings of the 40th International Conference on Machine Learning (ICML)}, volume 202 of \emph{Proceedings of Machine Learning Research}, pp.\  26619--26645, 2023.
\newblock URL \url{https://arxiv.org/abs/2302.01973}.

\bibitem[Quach et~al.(2024)Quach, Fisch, Schuster, Yala, Sohn, Jaakkola, and Barzilay]{quachconformal}
Victor Quach, Adam Fisch, Tal Schuster, Adam Yala, Jae~Ho Sohn, Tommi~S Jaakkola, and Regina Barzilay.
\newblock Conformal language modeling.
\newblock In \emph{International Conference on Learning Representations}, pp.\  11654--11681, 2024.

\bibitem[Ramalingam et~al.(2024)Ramalingam, Park, and Bastani]{ramalingam2024}
Ramya Ramalingam, Sangdon Park, and Osbert Bastani.
\newblock Uncertainty quantification for neurosymbolic programs via compositional conformal prediction, 2024.
\newblock URL \url{https://arxiv.org/abs/2405.15912}.

\bibitem[Reed et~al.(2022)Reed, DeVito, He, Ussery, and Ansel]{tracing}
James~K. Reed, Zachary DeVito, Horace He, Ansley Ussery, and Jason Ansel.
\newblock torch.fx: Practical program capture and transformation for deep learning in {Python}.
\newblock In \emph{Proceedings of Machine Learning and Systems (MLSys)}, volume~4, pp.\  638--651, 2022.
\newblock URL \url{https://arxiv.org/abs/2112.08429}.

\bibitem[Sandhu(1998)]{rbac}
Ravi~S. Sandhu.
\newblock Role-based access control.
\newblock In Marvin~V. Zelkowitz (ed.), \emph{Advances in Computers}, volume~46 of \emph{Advances in Computers}, pp.\  237--286. Elsevier, 1998.
\newblock \doi{10.1016/S0065-2458(08)60206-5}.
\newblock URL \url{https://www.sciencedirect.com/science/article/pii/S0065245808602065}.

\bibitem[Sandhu \& Samarati(1994)Sandhu and Samarati]{accesscontrol}
R.S. Sandhu and P.~Samarati.
\newblock Access control: principle and practice.
\newblock \emph{IEEE Communications Magazine}, 32\penalty0 (9):\penalty0 40--48, 1994.
\newblock \doi{10.1109/35.312842}.

\bibitem[Schick et~al.(2023)Schick, Dwivedi-Yu, Dess{\`\i}, Raileanu, Lomeli, Hambro, Zettlemoyer, Cancedda, and Scialom]{schick2023toolformer}
Timo Schick, Jane Dwivedi-Yu, Roberto Dess{\`\i}, Roberta Raileanu, Maria Lomeli, Eric Hambro, Luke Zettlemoyer, Nicola Cancedda, and Thomas Scialom.
\newblock {Toolformer}: Language models can teach themselves to use tools.
\newblock In \emph{Advances in Neural Information Processing Systems}, volume~36, pp.\  68539--68551, 2023.
\newblock \doi{10.52202/075280-2997}.

\bibitem[Surís et~al.(2023)Surís, Menon, and Vondrick]{surís2023vipergptvisualinferencepython}
Dídac Surís, Sachit Menon, and Carl Vondrick.
\newblock {ViperGPT}: Visual inference via {Python} execution for reasoning.
\newblock In \emph{Proceedings of the IEEE/CVF International Conference on Computer Vision (ICCV)}, pp.\  11854--11864, 2023.
\newblock \doi{10.1109/ICCV51070.2023.01092}.

\bibitem[Trivedi et~al.(2024)Trivedi, Khot, Hartmann, Manku, Dong, Li, Gupta, Sabharwal, and Balasubramanian]{trivedi2024appworldcontrollableworldapps}
Harsh Trivedi, Tushar Khot, Mareike Hartmann, Ruskin Manku, Vinty Dong, Edward Li, Shashank Gupta, Ashish Sabharwal, and Niranjan Balasubramanian.
\newblock {AppWorld}: A controllable world of apps and people for benchmarking interactive coding agents.
\newblock In \emph{Proceedings of the 62nd Annual Meeting of the Association for Computational Linguistics (Volume 1: Long Papers)}, pp.\  16022--16076, 2024.
\newblock \doi{10.18653/v1/2024.acl-long.850}.

\bibitem[Vovk et~al.(2005)Vovk, Gammerman, and Shafer]{vovk2005algorithmic}
Vladimir Vovk, Alexander Gammerman, and Glenn Shafer.
\newblock \emph{Algorithmic Learning in a Random World}, volume~29.
\newblock Springer, 2005.
\newblock \doi{10.1007/b106715}.

\bibitem[Wang et~al.(2024)Wang, Chen, Yuan, Zhang, Li, Peng, and Ji]{wang2024executablecodeactionselicit}
Xingyao Wang, Yangyi Chen, Lifan Yuan, Yizhe Zhang, Yunzhu Li, Hao Peng, and Heng Ji.
\newblock Executable code actions elicit better {LLM} agents.
\newblock In \emph{Proceedings of the 41st International Conference on Machine Learning (ICML)}, volume 235 of \emph{Proceedings of Machine Learning Research}, pp.\  50208--50232, 2024.

\bibitem[Wei et~al.(2022)Wei, Wang, Schuurmans, Bosma, Ichter, Xia, Chi, Le, and Zhou]{wei2022chainofthoughtpromptingelicitsreasoning}
Jason Wei, Xuezhi Wang, Dale Schuurmans, Maarten Bosma, Brian Ichter, Fei Xia, Ed~Chi, Quoc Le, and Denny Zhou.
\newblock Chain-of-thought prompting elicits reasoning in large language models.
\newblock In \emph{Advances in Neural Information Processing Systems}, volume~35, pp.\  24824--24837, 2022.
\newblock \doi{10.52202/068431-1800}.

\bibitem[Wu(2025)]{ptc}
Bin Wu.
\newblock Introducing advanced tool use on the {Claude Developer Platform}.
\newblock \url{https://www.anthropic.com/engineering/advanced-tool-use}, Nov 2025.
\newblock Accessed: 2026-03-31.

\bibitem[Yang et~al.(2024)Yang, Jimenez, Wettig, Lieret, Yao, Narasimhan, and Press]{yang2024sweagentagentcomputerinterfacesenable}
John Yang, Carlos~E. Jimenez, Alexander Wettig, Kilian Lieret, Shunyu Yao, Karthik Narasimhan, and Ofir Press.
\newblock {SWE-agent}: Agent-computer interfaces enable automated software engineering.
\newblock In \emph{Advances in Neural Information Processing Systems}, volume~37, pp.\  50528--50652, 2024.
\newblock \doi{10.52202/079017-1601}.

\bibitem[Yao et~al.(2023{\natexlab{a}})Yao, Yu, Zhao, Shafran, Griffiths, Cao, and Narasimhan]{yao2023treethoughtsdeliberateproblem}
Shunyu Yao, Dian Yu, Jeffrey Zhao, Izhak Shafran, Thomas~L. Griffiths, Yuan Cao, and Karthik Narasimhan.
\newblock Tree of thoughts: Deliberate problem solving with large language models.
\newblock In \emph{Advances in Neural Information Processing Systems}, volume~36, pp.\  11809--11822, 2023{\natexlab{a}}.
\newblock \doi{10.52202/075280-0517}.

\bibitem[Yao et~al.(2023{\natexlab{b}})Yao, Zhao, Yu, Du, Shafran, Narasimhan, and Cao]{yao2023reactsynergizingreasoningacting}
Shunyu Yao, Jeffrey Zhao, Dian Yu, Nan Du, Izhak Shafran, Karthik Narasimhan, and Yuan Cao.
\newblock {ReAct}: Synergizing reasoning and acting in language models.
\newblock In \emph{International Conference on Learning Representations}, 2023{\natexlab{b}}.
\newblock URL \url{https://openreview.net/forum?id=WE_vluYUL-X}.

\bibitem[Zhang et~al.(2025{\natexlab{a}})Zhang, Kraska, and Khattab]{zhang2025recursivelanguagemodels}
Alex~L. Zhang, Tim Kraska, and Omar Khattab.
\newblock Recursive language models, 2025{\natexlab{a}}.
\newblock URL \url{https://arxiv.org/abs/2512.24601}.

\bibitem[Zhang et~al.(2025{\natexlab{b}})Zhang, Li, and Bastani]{zhang2025}
Botong Zhang, Shuo Li, and Osbert Bastani.
\newblock Conformal structured prediction.
\newblock In \emph{International Conference on Learning Representations}, pp.\  8563--8583, 2025{\natexlab{b}}.

\end{thebibliography}
